\documentclass{article}
\usepackage{latexsym}
\usepackage{amsmath}
\usepackage{amssymb}
\usepackage{graphicx,color}

\makeatother

\newsavebox{\astrutbox}
\sbox{\astrutbox}{\rule[-5pt]{0pt}{20pt}}

\newcommand{\epsone}{\epsilon}
\newcommand{\epstwo}{\delta^2}
\newcommand{\epstwotwo}{\delta^4}

\newcommand{\intx}[1]{\partial^{-1} #1}
\newcommand{\rem}[1]{}
\newcommand{\Aco}{\tilde A}
\newcommand{\Bco}{\tilde B}

\newcommand{\aalpha}{a}

\newcommand{\sech}{{\rm sech}}

\rem{
\begin{figure}
\centerline{\includegraphics[width=3cm]{X.eps}}
\caption{\label{fig:X}}
\end{figure}
}
\rem{
\begin{figure}
     \vspace{5cm}
     \caption{An example figure with space for artwork}
     \label{sample-figure}
\end{figure}
}

\begin{document}

\title{\textbf{On asymptotically equivalent shallow water wave equations}}

\author{\textsf{H. R. Dullin$\,^{1,2}$, G. A. Gottwald$\,^3$ and
D. D. Holm$\,^{4,5}$}\\[6mm]
{\it{
$^1$ Mathematical Sciences, Loughborough University, Loughborough,
LE11 3TU, UK,}}
\\
{\small h.r.dullin@lboro.ac.uk}
\\
{\it{
$^2$ Fachbereich Physik, Universit\"at Bremen, D-28334 Bremen, Germany}}
\\
{\it{
$^3$ Mathematics and Statistics, University of Sydney, NSW 2006,
Australia,}}
\\
{\small gottwald@maths.usyd.edu.au}
\\
{\it{
$^4$ Theoretical Division and Center for Nonlinear Studies, Los Alamos
National Laboratory, }}
\\ 
{\it{
MS B284, Los Alamos, NM 87545, USA,}}\\ 
{\small email: dholm@lanl.gov}
\\
{\it{
$^5$ Mathematics Department, Imperial College of Science, Technology
and Medicine,}}
\\
{\it{
London SW7 2AZ, UK}}}

\date{03.07.2003}

\maketitle

\abstract{
The integrable 3rd-order Korteweg-de~Vries (KdV) equation emerges
uniquely at linear order in the asymptotic expansion for unidirectional
shallow water waves. However, at quadratic order, this asymptotic expansion
produces an entire {\it family} of shallow water wave equations that are
asymptotically equivalent to each other, under a group of nonlinear,
nonlocal, normal-form  transformations introduced by Kodama in combination
with the application of the Helmholtz-operator.
These Kodama-Helmholtz transformations are used to present connections
between shallow water waves, the integrable 5th-order Korteweg-de~Vries
equation, and a generalization of the Camassa-Holm (CH) equation that 
contains an
additional integrable case. The dispersion relation of the full water wave
problem and any equation in this family agree to 5th order. The travelling
wave solutions of the CH equation are shown to agree to 5th order with
the exact solution.
}

\setlength{\parskip}{1ex}

\section{Introduction}

We study the irrotational incompressible flow of a shallow layer of
inviscid fluid moving under the influence of gravity as well as
surface tension.  Previously \textsc{Dullin {\it{et al.}}} 
\cite{DGH[2001]} studied the case without surface tension, which in
the shallow water approximation leads to the Camassa-Holm equation
(CH). CH is the following 1+1 quadratically nonlinear equation for
unidirectional water waves with fluid velocity $u\,(x,t)$,
\begin{equation} \label{CH-BV}
       m_t  + c_0 m_x + u m_x + 2mu_x + \Gamma u_{xxx} = 0\,.
\end{equation}
Here $m=u-\alpha^{\,2}u_{xx}$ is a momentum variable, partial
derivatives are denoted by subscripts, the constants $\alpha^{\,2}$
and $\Gamma/c_0$ are squares of length scales and $c_0=\sqrt{gh}$ is
the linear wave speed for undisturbed water at rest at spatial
infinity, where $u$ and $m$ are taken to vanish. The limit
$\alpha^2\to0$ recovers the KdV equation,
\cite{KortewegdeVries[1895]}.

Equation (\ref{CH-BV}) was first derived by using asymptotic
expansions directly in the Hamiltonian for Euler's equations for
inviscid incompressible flow in the shallow water regime. It was
thereby shown to be bi-Hamiltonian and integrable by the inverse
scattering transform in the work of \textsc{Camassa \& Holm}
\cite{CH[1993]}.  Its periodic solutions were treated by \textsc{Alber
{\it{ et al}}} (see 
\cite{ACHM[1994]}, \cite{ACFHM[1999]} and references
therein). Before
\cite{CH[1993]}, families of integrable equations
similar to (\ref{CH-BV}) were known to be derivable in the general
context of hereditary symmetries by \textsc{Fokas \& Fuchsteiner} 
\cite{Fokas&Fuchssteiner}. However, equation (\ref{CH-BV}) was not
written explicitly, nor was it derived physically as a water wave
equation and its solution properties were not studied before
\cite{CH[1993]}. See \cite{Fuchssteiner} for an insightful discussion of
how the integrable shallow water equation (\ref{CH-BV}) relates to the
mathematical theory of hereditary symmetries.

Equation (\ref{CH-BV}) was recently rederived as a shallow water
equation by using asymptotic methods in three different approaches by
\textsc{Fokas \& Liu}
\cite{Fokas-Liu[1996]}, by \textsc{Dullin {\it{et al.}}}
\cite{DGH[2001]} and also by \textsc{Johnson}
\cite{Johnson2002}. These three derivations used
different variants of the method of asymptotic expansions for shallow
water waves in the absence of surface tension. We shall derive an
entire family of shallow water wave equations that are asymptotically
equivalent to equation (\ref{CH-BV}) at quadratic order in the shallow
water expansion parameters. This is one order beyond the linear
asymptotic expansion for the KdV equation. The asymptotically
equivalent shallow water wave equations at quadratic order in this
family are related amongst themselves by a continuous group of
nonlocal transformations of variables that was first introduced by
\textsc{Kodama}
\cite{Kodama[1985&1987]}.

\textsc{Dullin {\it{et al.}}} \cite{DGH[2001]} focused on the integrability of  equation
(\ref{CH-BV}) and its spectral properties. Its derivation from Euler's
equation in the case without surface tension was briefly
described. Here we present the necessary details of this calculation.
The present derivation also adds surface tension.  In view of the many
papers which have appeared recently on weakly nonlinear shallow water
models (i.e. \cite{Fokas[1995],Johnson2002}) we see our contribution
to be the following. By combining a nonlocal Kodama transformation
with the application of the smoothing Helmholtz operator, we derive
several integrable water wave equations in the same asymptotically
equivalent family. This family includes the CH equation (\ref{CH-BV}),
the fifth-order Korteweg de Vries equation (KdV5) and the equation of
\textsc{Degasperis \& Procesi} \cite{DP[1999]} which was recently discovered to be integrable in
\cite{DHH2002}. All these integrable equations are then explicitly related to each
other, again by means of a Kodama transformation. We clarify the
differences among the previous derivations of these equations. The
equations are discussed with respect to their linear dispersion
properties.

In the context of water waves subject to surface tension, interest has
recently focused on the KdV5 equation and its solitary wave
solutions. See \cite{DiasKharif[1999]} for a review.  For Bond numbers
$0<\sigma<1/3$ it has been shown that these solutions are not true
solitary waves which decay to zero at spatial infinity but instead
they are {\it{generalized solitary waves}} which are characterized by
exponentially small ripples on their tail. See for example
\cite{Beale[1991]} for more explanation.  It has been shown by \textsc{Lombardi}
\cite{Lombardi[2000]} that these ripples are exponentially small in
terms of $F-1$ where $F=c/c_0$ is the Froude number. Numerical
experiments by \textsc{Champneys} \cite{Champneys} suggest that in the
full nonlinear water wave problem there are no real solitary waves
bifurcating for Bond numbers $0<\sigma<1/3$.  For Bond numbers larger
than $1/3$, one obtains depressions with negative velocity, rather
than elevations with positive velocity.

One may ask whether yet another model equation is needed, if the more
rigorous, exact, or numerical results described above are already
available.
Or in more general terms: Which is preferred? An exact solution of an
approximate model equation, or an approximate solution of an exact
equation? Our point of view is: If the added cost is small, why not
take both? It might be useful to have an equation that gives, e.g.,
more accurate travelling waves than KdV, without the need to go to
much more elaborate models.  Although we shall obtain less information
and accuracy than the sophisticated, beyond-all-order methods, we
shall also pay less.  Thus, one can improve the description of the
shape and speed of the travelling wave without having to resort to
these more complicated models.  That they are still only an
approximation to the true solution is, of course, taken for granted.

Our inclusion of surface tension has a similar motivation.  Although
the equation (\ref{peyresq}) that we shall derive has some drawbacks
concerning the global properties of its dispersion relation for large
$k$, it still gives improved descriptions for small $k$ and small
$\sigma$.  Moreover, the improved solutions are easily obtained and
analyzed.

\noindent{\bf Outline}
Section \ref{sec:eta} recalls the standard dynamics for the shallow
water wave elevation following \textsc{Whitham}
\cite{Whitham[1974]}. We then use an approach based on the Kodama
transformation to derive equation (\ref{CH-BV}) with surface tension
in section \ref{sec:inteq}.  Section \ref{relation} explores the
transformations employed to derive water wave equations. We discuss
the class of equations which may be related to each other via such
transformation, and are, hence, asymptotically equivalent. As
examples, we discuss the relations of equation (\ref{CH-BV}) to KdV
and other integrable equations. We particularly discuss the relations
to KdV5, the fifth-order integrable equation in the KdV hierarchy, and
to another integrable nonlinear equation recently proposed by
\textsc{Degasperis \& Procesi} \cite{DP[1999]} and discovered to be
integrable in \cite{DHH2002}. Section \ref{sec:dis} compares the
dispersion relation of (\ref{CH-BV}) with that of the full water wave
equation.  Finally section \ref{sec:tw} shows that the travelling wave
solutions of (\ref{CH-BV}) agree to fourth order with the exact
travelling waves solutions of the full water wave problem.

\section{The $\boldsymbol{\eta}$ equation}
\label{sec:eta}
Our derivation of equation (\ref{CH-BV}) proceeds from the physical
shallow water system along the lines of \textsc{Whitham} \cite{Whitham[1974]}. Consider
water of depth $h = h_0 + \eta(x,t)$ where $z=-h_0$ is the flat bottom
and $h_0$ the mean depth, so that $z = 0$ at the free surface in
equilibrium. Denote by $u_h$ and $u_v$ the horizontal and vertical
velocity components, respectively.  The $z$-momentum equation is
\begin{equation} \label{eqn:Dw}
      \frac{Du_v}{Dt} = -g - \frac{1}{\rho} \partial_z p
       \quad\hbox{with}\quad
        p = \tilde\sigma \frac{h_{xx}}{({1 + h_x^2})^{3/2}}
      \,,
\end{equation}
where $g$ is the constant of gravity and $\tilde \sigma$ is the
surface tension. At the free surface, the boundary condition is
\begin{equation} \label{eqn:Deta}
      \frac{D\eta}{Dt} = u_v \,  \quad \mbox{where } z =  \eta \,.
\end{equation}
The irrotational velocity
$\mathbf{u}(x,z,t)=\nabla\varphi$ has horizontal and vertical velocity
components $u_h=\varphi_x$ and $u_v=\varphi_z$. The velocity potential
$\varphi$ must satisfy Laplace's equation in  the interior.
Eq.~(\ref{eqn:Deta}) yields the kinematic boundary
condition at the free surface,
\[
       \eta_t + \varphi_x \eta_x = \varphi_z\,.
\]
Eq. (\ref{eqn:Dw}) can now be integrated to yield
the dynamic boundary condition,
\[
      \varphi_t + \frac12 (\varphi_x^2 + \varphi_z^2)
      = -g h - \frac{1}{\rho} p \,.
\]
The equations for a fluid are non-dimensionalized by
introducing $x=l_x x'$, $z=h_0 z'$, $t=(l_x/c_0) t'$, $\eta=a \eta'$
and $\varphi=(g l_x a/c_0)\varphi'$, where $c_0 =
\sqrt{g h_0}$. Being interested in weakly nonlinear, small-amplitude
waves in a shallow water environment, we introduce the small parameters
$\epsone=a/h_0$ and $\epstwo=(h_0/l_x)^2$ where $\epsone \ge
\epstwo > \epsone^2 \ge
\epsone\epstwo
\ge \epstwotwo$.
When we talk about the linear approximation we lump together the terms
or order $\epsone$ and $\epstwo$, because they are allowed to be of
the same order. Similarly, quadratic order means quadratic in $\epsone$
and $\epstwo$, hence includes terms of order
$\epsone^2, \epsone\epstwo, \epstwotwo$. Contrary to this convention
the traditional designations 3rd or 5th order KdV refer to the power of $\delta$, and
hence also the maximal number of derivatives in the linear terms.
Upon omitting the primes and expanding the pressure term up to order
$\epsone^2\epstwo$, the Euler equations and the boundary conditions at
the free surface and at the bottom are expressed as,
\begin{eqnarray}
\label{euler2}
\epstwo\varphi_{xx}+\varphi_{zz}=0
\qquad &{\rm{in}}& \qquad
-1<z<\epsone \eta \\ \label{eqn:EBC1}
\eta_t+\epsone \varphi_x \eta_x
-\frac{1}{\epstwo}\varphi_z = 0
\qquad &{\rm{at}}& \qquad
z=\epsone \eta\\
\eta+\varphi_t+\frac{1}{2}(\epsone\varphi_x^2
   +\frac{\epsone}{\epstwo}\varphi_z^2)
-\sigma \epstwo \eta_{xx}
   = 0   
\qquad &{\rm{at}}& \qquad
z=\epsone \eta\\
\varphi_z=0
\qquad &{\rm{at}}& \qquad
z=-1 \; ,
\end{eqnarray}
where $\sigma = \tilde \sigma / (h_0 \rho c_0^2)$ is the dimensionless
Bond number.
The ordering of $\epsone$ and $\epstwo$ is as specified, provided
that ${\cal O}(\sigma) = 1$. 

It is well known that variable transformations of the water wave
problem yield the following decoupled equation for the elevation, $\eta$,
\begin{eqnarray}
    \nonumber  \eta_t
    + \eta_x
    + \frac{3}{2}\epsone\eta \eta_x
    + \frac{1}{6}\epstwo (1 - 3\sigma) \eta_{xxx}
    - \frac{3}{8}\epsone^2 \eta^2\eta_x
    + \epsone\epstwo
      \Big( \frac{1}{24} (23 + 15\sigma)\eta_x\eta_{xx}&& \\
       +\ \frac{1}{12} (5 - 3\sigma)\eta\eta_{xxx} \Big)
    + \epstwotwo\frac{1}{360} (19 - 30\sigma -
45\sigma^2) \eta_{xxxxx} =
0. && \qquad \label{asy1}
\end{eqnarray}
The derivation up to this order appears, for example, in
\cite{MarchantSmyth[1990]}, or more
recently in \cite{Johnson2002} without surface tension.

\section{Transformation to an integrable equation}
\label{sec:inteq}

Before embarking on its derivation, we shall discuss the transformation
properties of equation (\ref{CH-BV}). First, it is reversible, i.e.,
it is invariant under the discrete transformation $u(x,t)\to-u(x,-t)$.
Equation (\ref{CH-BV}) is also Galilean {\it covariant}. That is, it
keeps its form under transformations to an arbitrarily moving
reference frame. This includes covariance under transforming to a
uniformly moving Galilean frame.  However, equation (\ref{CH-BV}) is
not Galilean {\it invariant}, even assuming that the momentum, $m$,
Galileo-transforms in the same way as velocity, $u$. In fact, equation
(\ref{CH-BV}) transforms under
\begin{equation}\label{xtrans}
\hspace{-3mm}
t\to t+t_0\,, \quad
x\to x+x_0+ct\,, \quad
u\to u+c+u_0\,, \quad
m\to m+c+u_0\,,
\end{equation}
to the form
\begin{equation}\label{b-family-xtrans}
m_t+\, um_x+2u_x\,m
+(c_0+u_0)m_x
+2u_x(c+u_0)
+ \Gamma u_{xxx}
=0
\,.
\end{equation}

Thus, equation (\ref{CH-BV}) is invariant under space and time
translations (constants $x_0$ and $t_0$), covariant under Galilean
transforms (constant $c$), and acquires linear dispersion terms under
velocity shifts (constant $u_0$).  The dispersive term $u_0m_x$
introduced by the constant velocity shift $u_0\ne0$ breaks the
reversibility of equation (\ref{CH-BV}).

Under scaling transformations of $x$, $t$ and $u$, the coefficients of
equation (\ref{CH-BV}) may be changed.  However, such scaling leaves
the following coefficient ratios invariant,
\begin{eqnarray}
C(u_x u_{xx}) &:&  C( u u_{xxx}) = 2 : 1\label{C2} \,,\\
C(u_{xxt}) C( u u_x) &:& C( u u_{xxx} ) C( u_t) = 3 : 1 \,,
\label{C3}
\end{eqnarray}
where $C(f)$ stands for the coefficient of $f$ in the scaled
equation. It is pertinent to mention that the above ratios are crucial
in the integrability of equation (\ref{CH-BV}). See \cite{CH[1993]} and
\cite{DGH[2001]} for discussions of this point.

Equation (\ref{CH-BV}) will emerge as being asymptotically equivalent
to equation (\ref{asy1}) after two steps. First, we shall
perform a near-identity transformation
\begin{eqnarray}
\eta=\eta(u) = u + \epsone f[u] + \epstwo g[u]
\,,
\label{near-id-trans}
\end{eqnarray}
relating the wave elevation and a ``velocity-like'' quantity, $u$. One
should not regard $u$ as the original fluid velocity, because we will
use a non-local term in $f$ which is difficult to interpret in this
context. Instead, we consider $u$ as an auxiliary quantity in which
the equation becomes particularly simple. The quantity $u$ agrees at
leading order with the fluid velocity and it transforms as a velocity
under time reversal, spatial reflection and Galilean boosts.  To
obtain the physically more meaningful quantity $\eta$ one has to
transform back, see below. The functionals $f$ and $g$ in the
transformation (\ref{near-id-trans}) are to be chosen so that they
generate the terms proportional to $u u_x$, $u_x u_{xx}$, $u u_{xxx}$
and $u_{xxx}$ in equation (\ref{CH-BV}). Afterwards, we shall apply
the Helmholtz operator $H = 1 - \nu \epstwo\partial_x^2$, which
generates the $u_{xxt}$ term.  As in \cite{Kodama[1985&1987]} the
functional $g[u]$ is proportional to $u_{xx}$ and $f[u]$ is a linear
combination of $u^2$ and a non-local term $u_x \partial^{-1}$, where
$\partial^{-1}$ denotes integration over $x$.  Thus, including the
parameter $\nu$, there are four coefficients in this {\em
Kodama-Helmholtz} transformation (KH-transformation).  These shall be
chosen so
that equation (\ref{CH-BV}) emerges, after a rescaling of $u$, $x$ and
$t$.

With these choices, (\ref{near-id-trans}) becomes the Kodama
transformation, which depends on three parameters $\alpha_1$,
$\alpha_2$ and $\beta$,
\begin{eqnarray}
\label{eta} \label{Kodama}
      \eta
    = \eta(u)
    = u
    + \epsone(\alpha_1u^2+\alpha_2u_x\intx{u})
    + \epstwo\beta u_{xx}\; .
\end{eqnarray}
Terms of degree $n$ in the expansion parameters $\epsone$ and
$\epstwo$ start contributing at degree $n+1$ in the transformed
equation. Therefore, no terms of quadratic order in $\epsilon$ and
$\delta^2$ are needed in the transformation. Inserting the
Kodama transformation (\ref{eta}) into equation (\ref{asy1}) for the
height field $\eta$ leads to the following terms in asymptotic order.
\begin{eqnarray}
\nonumber
{\cal{O}}(1)&:&\quad
      u_t + u_x\\ \nonumber
{\cal{O}}(\epsone)&:&\quad
      2\alpha_1 u u_t
     +2\alpha_1 u u_x
     +\alpha_2
     ( u_{xt}\intx{u}
      +u_{xx}\intx{u}
      +u_x\intx{u_t}
      +u u_x
     )
     +\frac{3}{2}uu_x\\
{\cal{O}}(\epstwo)&:&\quad    \label{uxxtsub}
      \beta u_{xxt}
     +u_{xxx}( \beta + \frac16 - \frac12 \sigma) \\ \nonumber
{\cal{O}}(\epsone^2)&:&\quad
      \frac{9}{2}\alpha_1 u^2u_x
     +\frac{3}{2}\alpha_2
     ( u^2u_x
      +uu_{xx}\intx{u}
      +u_x^2\intx{u} )
     -\frac{3}{8}u^2u_x\\ \nonumber
{\cal{O}}(\epsone\epstwo)&:&\quad
     ( \frac{23}{24} + \frac58 \sigma
      +\frac13 ( 3\alpha_1 + 2\alpha_2 )(1-3\sigma)
      +\frac{3}{2}\beta
       )   u_xu_{xx} + \\ && \nonumber
    +( \frac{5}{12}   -\frac14 \sigma
      +\frac{1}{6}(2\alpha_1+3\alpha_2)(1-3\sigma)
      +\frac{3}{2}\beta )    uu_{xxx} 
    + \frac16 \alpha_2 (1 - 3\sigma) u_{xxxx}\intx{u} \\ \nonumber
{\cal{O}}(\epstwotwo)&:&\quad
(\frac16 \beta_2(1-3\sigma) + \frac{1}{360}(19 -
30\sigma -
45\sigma^2) ) u_{xxxxx}
\;.
\end{eqnarray}
As before, we expand the time derivatives to linear order as
\begin{eqnarray}
u_t
&=&
-u_x
-\frac{3}{2}\epsone uu_{x}
-\frac{1}{6}\epstwo (1 - 3\sigma) u_{xxx}
\; , \\ \nonumber
u_{xt}
&=&
-u_{xx}
-\frac{3}{2}\epsone u_{x}^2
-\frac{3}{2}\epsone uu_{xx}
-\frac{1}{6}\epstwo (1 - 3\sigma) u_{xxxx}
\; , \\
u_{xxt} \label{uxxtis}
&=&
-u_{xxx}
-\frac{9}{2}\epsone u_xu_{xx}
-\frac{3}{2}\epsone uu_{xxx}
\;.
\end{eqnarray}
This expansion generates higher order terms, leading to
\begin{eqnarray}
{\cal{O}}(1)&:&\quad
    u_t
+u_x
\nonumber \\
{\cal{O}}(\epsone)&:&\quad
\frac{3}{2}uu_x
\nonumber  \\
{\cal{O}}(\epstwo)&:&\quad
\frac{1}{6}(1-3\sigma)u_{xxx}
\label{delta-squared-terms} \\
{\cal{O}}(\epsone^2)&:&\quad
( \frac{3}{2}\alpha_1
     +\frac{3}{4}\alpha_2
     -\frac{3}{8} )  u^2u_x
\label{beforeabeq} \\ \label{acobcoeq}
{\cal{O}}(\epsone\epstwo)&:&\quad
\Aco u_xu_{xx}
+ \Bco  uu_{xxx} \\
{\cal{O}}(\epstwotwo)&:&\quad
     \frac{1}{360}( 19 - 30\sigma - 45 \sigma^2 )
u_{xxxxx}
\;.
\label{asy_sys}
\end{eqnarray}
where in (\ref{acobcoeq}) we defined
\[
\Aco = \frac{23}{24} + \frac58 \sigma
     +\frac12 (2\alpha_1+\alpha_2)(1-3\sigma)
     -3 \beta \quad \mbox{ and } \quad
\Bco = \frac{5}{12} -\frac14 \sigma
     +\frac{1}{2}\alpha_2 ( 1 - 3\sigma) \,.
\]
The first step of the derivation is now complete.  In the second step,
applying the Helmholtz operator $H = 1 - \nu\epstwo \partial_x^2$
introduces the coefficient $\nu$ and creates terms with two more $x$
derivatives. However, the terms of order ${\cal{O}}(\epsone^2)$ are
left unchanged. These terms are proportional to $u^2 u_x$ and they
must vanish for equation (\ref{CH-BV}) to emerge. Thus, the
application of the Helmholtz operator restores the needed $u_{xxt}$
term that had previously been eliminated.  Alternatively, the same
equation (\ref{CH-BV}) can be obtained by splitting the time
derivative, that is, by partially substituting the time derivative
$u_{xxt}$ in (\ref{uxxtsub}) by its asymptotic approximations
(\ref{uxxtis}), along the lines of \textsc{Benjamin {\it{et al.}}} 
\cite{BBM} in their study of the BBM equation.

The order ${\cal{O}}(\epsone^2)$ coefficient in expression
(\ref{beforeabeq}) will vanish, provided the parameters $\alpha_1$ and
$\alpha_2$ are chosen to satisfy
\begin{eqnarray}
\label{kod1}
    4\alpha_1 + 2\alpha_2 = 1
\; .
\end{eqnarray}
The order ${\cal O}(\epstwotwo)$ terms receive an additional contribution
that arises from applying the Helmholtz operator $H = 1 - \nu\epstwo
\partial_x^2$ to the terms of order ${\cal O}(\epstwo)$. The entire
combination at order ${\cal O}(\epstwotwo)$ must vanish, for  the final
equation to possess no fifth-order derivative term,
$u_{xxxxx}$. This requirement determines $\nu$ as
\begin{equation} \label{nu}
      \nu = \frac{1}{60} \frac{19 - 30\sigma -
45\sigma^2}{1-3\sigma}
\,.
\end{equation}
In what follows, we shall consider the coefficient $\nu$ to be given
by this function of surface tension, $\sigma$.  Note: this removal of
the highest order term was made possible by introducing the additional
parameter $\nu$ via the Helmholtz operator.  The remaining terms
containing free parameters
$\alpha_2$ and $\beta$ are of order $\epsone\epstwo$ and they
combine additively as
\[
      (\Aco  - \frac92 \nu ) u_x u_{xx} + (\Bco -
\frac32 \nu) u u_{xxx} \,.
\]
To ensure equivalence to (\ref{CH-BV}) except for scaling we need the
relative coefficients to appear in the ratio (\ref{C2}), so that
\begin{equation} \label{eqn:twotoone}
(\Aco - \frac92\nu) : (\Bco-\frac32\nu) = 2:1 \,.
\end{equation}
In addition we also need to satisfy (\ref{C3}), so that
\[
        \frac32 \nu : (\Bco - \frac32 \nu)  = 3:1 \,.
\]
These two conditions imply $\Bco = 2\nu$ and $\Aco = 11\nu/2$.  As a
result we finally obtain the equation
\begin{equation} 
       u_t - \nu \epstwo u_{xxt} + u_x + \frac32
\epsone u u_x -
         \frac{1}{2}\epsone\epstwo \nu ( u u_{xxx}
+ 2 u_x u_{xx})
+ \epstwo (\frac16 - \nu - \frac12 \sigma )u_{xxx}
= 0\,,
\end{equation}
which can be rewritten in terms of $m = u -
\nu\epstwo u_{xx}$ as
\begin{equation} \label{final}
       m_t  + m_x + \frac{\epsone}{2} ( u m_x + 2mu_x
)
+ \epstwo (\frac16  - \frac12 \sigma )u_{xxx} =
0\,.
\end{equation}
Thus, the coefficients in the Kodama transformation (\ref{Kodama})
that yield this equation are,
\begin{eqnarray} 
    \alpha_1 &=&  \frac{7}{20} - \sigma\frac{1}{5}
\frac{2-3\sigma}{(1-3\sigma)^2} \,,\\
    \alpha_2 &=&  -\frac{1}{5} + \sigma\frac{2}{5}
\frac{2-3\sigma}{(1-3\sigma)^2} \,,\\
    \beta    &=&  \frac{1}{30} - \sigma\frac{1}{30}
\frac{17-30\sigma}{1-3\sigma} \,.
\end{eqnarray}

Returning to physical variables, where $u$ and
$\varphi_x$ have  units of $ga/c_0 = c_0 a/h_0$, followed by an additional
scaling of $u\to2u$, gives equation (\ref{final}) the canonical CH
form,
\begin{equation} \label{peyresq}
       m_t  + c_0 m_x + u m_x + 2mu_x  +
\Gamma u_{xxx} = 0\,,
\end{equation}
in which $m = u - \nu h^2 u_{xx}$ and $\Gamma = {c_0h^2}(1-3\sigma)/6$.
The parameters $\alpha^2$ and $\Gamma$ in CH (\ref{CH-BV}) are
now understood in terms of physical variables as
\begin{eqnarray} \label{alpha-gamma}
    \alpha^2 = \nu h^2  = h^2 \frac{1}{60} \frac{19 -
30\sigma -
45\sigma^2}{1-3\sigma}
\,,\quad
    \Gamma  = \frac{c_0h^2}{6}(1-3\sigma)\,.
\end{eqnarray}
The parameter $\Gamma$ changes sign when the Bond number $\sigma$
crosses the critical value, $1/3$.
For later reference, we record the value of $\sigma >0 $ for which
$\alpha^2$ vanishes, as
\begin{equation} \label{sigmacrit}
      \sigma_\alpha = -\frac13 + \frac{2}{15}\sqrt{30}
\simeq 0.39696 > \frac13
\,.
\end{equation}
%

In the special case $c_0=\Gamma=0$ equation (\ref{peyresq}) is called
``the peakon equation.'' This equation supports {\it peakons} as
solitary wave solutions whose derivative is discontinuous at the
extremum. These solutions were introduced and discussed by
\textsc{Camassa \& Holm}
\cite{CH[1993]}. The peakon equation has many exceptional mathematical
properties that arise from its interpretation as geodesic motion in
the Euler-Poincar\'e variational framework, as explained in
\cite{HMR[1998]}. However, the peakon equation cannot be derived as a
water wave equation in a weakly nonlinear shallow approximation from
the Euler equation by the present technique. This is because neither
a Galilean transformation, nor an appropriate splitting can eliminate
both of the linear dispersive terms in equation (\ref{peyresq})
simultaneously. One is always left with a residual linear dispersion.

\textsc{Johnson} \cite{Johnson2002} has recently derived the dispersive CH equation
(\ref{peyresq}) as a shallow water wave equation by using the same
asymptotic expansion. However, two key steps in the asymptotic
derivation are addressed in quite different ways in \cite{Johnson2002}
and here.  Firstly, the ratios of the coefficients of the equation
must be adjusted to ensure integrability. In
\cite{Johnson2002}, this is achieved by using the height at which the
velocity potential is evaluated as a free parameter. Instead of the
height, we use the free parameters in the Kodama transformation
(\ref{Kodama}) to obtain the desired ratios.

Secondly, the fifth-derivative term $u_{xxxxx}$ of order ${\cal
O}(\delta^4)$ must be removed. The present approach allows the value
of the free parameter $\nu$ in the Helmholtz operator to be chosen to
cancel out this term. In \cite{Johnson2002}, this term is simply
omitted.
In Section 4 we shall use our approach to show that the CH equation is
asymptotically equivalent to the KdV5 equation, which involves this
fifth-order derivative.

In order to compare predictions in terms of physically measurable
quantities, the solutions for the 
velocity-like variable 
$u$ must be
transformed back to the elevation field $\eta$ by using (\ref{Kodama}).
However, the derivation not only used the transformation (\ref{Kodama}),
it also involved applying the Helmholtz operator.  Therefore,
one should verify that it is sufficient to simply invert (\ref{Kodama}).
Fortunately, when the inverse transformation $u=u(\eta)$ of the same form
as (\ref{Kodama}) with $u$ and $\eta$ interchanged is substituted into the
final equation (\ref{final}), we find that the coefficients just reverse
their signs. We conclude that (\ref{CH-BV}) is equivalent to the shallow
water wave equation (\ref{asy1}) up to and including terms of order ${\cal
O}(\epstwotwo)$.

\section{The Kodama-Helmholtz transformation}
\label{relation}

In the previous section, we transformed the shallow water wave
equation (\ref{asy1}) into the CH-equation (\ref{CH-BV}) by means of a
Kodama transformation (\ref{Kodama}) and an application of the
Helmholtz-operator. Now we describe the class of equations that can be
derived from the shallow water wave equation (\ref{asy1}) by any such
sequence of transformations, which we will refer to as
KH-transformation. This class contains (at least) four integrable
equations, one of which is the CH-equation.  Moreover, all equations
in this class are not only related to (\ref{asy1}); they are also
related to each other by such transformations and are therefore
asymptotically equivalent.

Similar transformations have been used by \textsc{Fokas \& Liu}
\cite{Fokas-Liu[1996]} that included an additional term of the form $x
u_t$, but no Helmholtz-operator. In this case, the class of equations
even includes the (3rd order) KdV equation.  Unfortunately, though,
this term is not uniformly bounded, so we shall decline to use it. Its
unboundedness is a problem when transforming travelling wave solutions
which move towards $x=\pm\infty$. Moreover, use of the term $x u_t$
changes the dispersion relation.

\subsection{Invariance of the dispersion relation}
\label{subsec:disprel}

The Kodama transformation (\ref{Kodama}) does not change the dispersion
relation. To see this, one may observe that only terms linear in $\eta$ or
its derivatives produce linear terms in the transformed equation.
Similarly, applying (\ref{Kodama}) to nonlinear terms in an equation
produces only nonlinear terms in the transformed equation. Therefore, in
proving invariance of the linear dispersion relation under (\ref{Kodama}),
we may restrict to a transformation $\eta = u +\varepsilon L(u)$ in which
$L$ is a linear differential operator with constant coefficients and the
linear equation to be transformed is $\eta_t = M(\eta)$.  To first order,
we then have $u_t = M(u)$ and the full transformation gives
\[
        u_t + \varepsilon L(u_t) = M(u + \varepsilon L(u))\,.
\]
Now the first order equation may be used to eliminate the time derivatives
which are not of order zero, thereby yielding
\[
     u_t + \varepsilon L(M(u)) = M(u) + \varepsilon M(L(u))
\,.
\]
If $M$ and $L$ commute, as they always do when they have constant
coefficients, the final answer is $u_t = M(u)$. Consequently, the Kodama
transformation (\ref{Kodama}) leaves a linear equation unchanged.
Note that including the $x u_t$ term in the transformation would in general
cause the operators to no longer commute, so that the dispersion relation 
would be changed, as previously claimed.
%

The second step of the transformation is the application of the
Helmholtz operator $H = 1 - \nu \epstwo \partial_x^2$.
As we have just seen, the Kodama transformation leaves the linear part of
the equation unchanged.  Applying the Helmholtz operator to an equation
does change the linear part, but it still leaves the dispersion relation
unchanged. To see this, let the linear part of the equation be given by
$u_t = M(u)$.  The new equation is $H(u_t) = H(M(u))$. If $H$ and $M$ are
linear with constant coefficients this gives $H(u)_t = M(H(u))$ so
that with the definition $m = H(u)$ we obtain $m_t = M(m)$, which has
the same dispersion relation. This does not hold, however, if we
truncate higher order terms in $H(M(u))$.  If we truncate, then the
dispersion relation will agree up to the order of truncation. For
example, the dispersion relation for (\ref{CH-BV}) is a rational
function, which differs from the polynomial dispersion relation
obtained from (\ref{asy1}). However, by the above argument the two
dispersion relations agree up to the desired order.

\subsection{Range of the KH-transformation}

We now investigate which equations may be transformed into each
other by a  KH-transformation.
The class of equations that (under some additional conditions to
be derived) may be transformed into each other by
KH-transformations is given by
\begin{equation}
\label{FG}
F(u_t,u_x,u_{xxx},u_{xxt},u_{xxxxx},uu_x) +
G(u^2u_x,uu_{xxx},u_xu_{xx}) = 0\,,
\end{equation}
where $F$ and $G$ are linear and the coefficients of $u_t$, $u_x$,
$uu_x$, $u_{xxx}$ are nonzero. This means that each equation has a
KdV-kernel. In addition, we assume that the terms are ordered as in the
previous section: Every $u$ has weight $\epsilon$ and every $x$
(or $t$) derivative has weight $\delta$. However, for simplicity, we
do not explicitly display the weights in the following, even though they
are used for truncation at the usual order.
The general procedure consists of two parts. In the first part the
coefficients in $F$ are normalised in three steps. In the second
part the Kodama transformation is used to adjust the terms in $G$.
The details of the calculation are similar to those of the previous
section and are therefore not given.

Equations are considered to be equivalent when they differ only by a
transformation of the form
\begin{equation} \label{txuTraf}
(t,x,u) \to (\tau t, \xi x - \kappa t, \nu u)\,.
\end{equation}
Scaling of $u$, $t$, and the equation allow one to set the coefficients of
$u_x$, $u_t$, and $uu_x$ to arbitrary non-zero values. After such a
scaling, the equation takes the form,
\begin{equation} \label{step1}
u_t + u_x + uu_x + f_3 u_{xxx} + f_4 u_{xxt} +  f_5 u_{xxxxx} +
g_1 u^2 u_x + g_2 uu_{xxx} + g_3 u_x u_{xx} = 0\,.
\end{equation}
The next step is to remove the $u_{xxt}$ term by eliminating
the $t$ derivatives using the equation itself.
This operation produces an equation of the form,
\begin{eqnarray} \nonumber
&& u_t + u_x +  uu_x +( f_3 -f_4) u_{xxx} +  (f_5 -f_4(f_3-f_4) ) 
u_{xxxxx} + \\
&& \qquad g_1 u^2 u_x + (g_2-f_4) uu_{xxx} + (g_3-3f_4) u_x u_{xx} = 
0\,. \label{step2}
\end{eqnarray}
In the third step, the coefficients of $u_{xxx}$ and $u_{xxxxx}$ are
normalised using (\ref{txuTraf}) again, 
while keeping the coefficients of $u_t$, $u_x$, $uu_x$ all fixed at unity. 
First $\xi$ is used to normalize the ratio between the
coefficients of $u_{xxx}$ and $u_{xxxxx}$ to $\pm 1$, assuming that both
are non-zero.  
The relative sign of these coefficients cannot be changed.
Dividing the equation by $\tau$ allows both coefficients to be set to
unit magnitude. 
In order to keep the other coefficients in the linear
function $F$ equal to unity, 
the special transformation of the form (\ref{txuTraf}) reads
$(t,x,u) \to (\tau t, \xi x  + (\tau - \xi)  t, \xi u /\tau ) $.
The above three steps (transformation (\ref{txuTraf}) to find (\ref{step1}),
elimination of $u_{xxt}$ to find (\ref{step2}) and
again transformation (\ref{txuTraf}) to find (\ref{step3}))
can be performed for any equation in the class (\ref{FG}).
Therefore, we shall only be concerned with the equivalence
under Kodama transformation of equations of the form
\begin{equation} \label{step3}
u_t + u_x + \tilde f_6 uu_x + \tilde f_3 u_{xxx} + \tilde f_5 u_{xxxxx} +
\tilde g_1 u^2 u_x + \tilde g_2 uu_{xxx} + \tilde g_3 u_x u_{xx} = 0\,,
\end{equation}
where $\tilde f_6 = 1$, $\tilde f_3 = 1$, or 0, and $\tilde f_5 = \pm1$, or
0. $\tilde f_3 = 0$ appears when $f_3 = f_4$ in (\ref{step1}).
The parameter $\tilde f_6$ is kept in the notation to trace the 
influence of the $uu_x$ term. 
The terms in the linear function $F$ are unchanged by the Kodama
transformation because such linear terms are unchanged by a near identity
transformation and $uu_x$ is the lowest order nonlinearity.
Therefore two equations can only be equivalent if their
normalised coefficient $\tilde f_5$ agrees. Recall that $u_{xxt}$ has
already been eliminated, so that if, e.g., the above steps are applied
to the CH-equation, the resulting equation does possess a non-zero
fifth derivative term.

The terms in the linear function $G$ in (\ref{FG}) can be adjusted by
the Kodama transformation. This generates coefficients,
\begin{eqnarray}
\hat g_1 &=& g_1 + \tilde f_6( 2 \alpha_1  + \alpha_2 )/2
\,,\nonumber \\
\hat g_2 &=& g_2 + 3 \alpha_2 \tilde f_3
\,, \\
\hat g_3 &=& g_3 + 3 \tilde f_3 ( 2 \alpha_1+\alpha_2)- 2\beta \tilde f_6
\,. \nonumber
\end{eqnarray}
The $\hat g_i$ are linear in the coefficients
$\alpha_1$, $\alpha_2$, and $\beta$ of the Kodama
transformation. A sufficient
condition for the solvability of this linear
system is
$
    \tilde f_3 \tilde f_6^2 \not = 0\,.
$
Hence, if the coefficients of $u_{xxx}$ and $uu_x$ in (\ref{step3}) are
non-zero, then any value of $\hat g_i$ can be achieved by some Kodama
transformation.
Since the coefficient of $uu_x$ in (\ref{FG}) is assumed to be non-zero
the only additional condition is $\tilde f_3 = f_3 - f_4 \not = 0$ in
(\ref{step1}). Note that this condition was already necessary in order to
obtain (\ref{step3}).
For example, any equation with positive relative sign of $\tilde f_3$ and
$\tilde f_5$ in (\ref{step3}) is equivalent to the water wave equation
(\ref{asy1}) with $\sigma = 0$. 
In particular, this shows that the integrable KdV5 can be obtained, see
\cite{LiZhiSibgatullin[1997]}.  Similarly, different higher-order, but
non-integrable, extensions of the KdV equation can also be obtained,
e.g., those introduced by \textsc{Champneys \& Groves} \cite{ChampneysGroves}. Notice that the
relative sign of $\tilde f_3$ and $\tilde f_5$ in (\ref{step3}) is
negative for $1/3 <
\sigma <\sigma_\alpha$, see (\ref{sigmacrit}). The condition $\tilde f_3=0$
appears for $\sigma = 1/3$, while $\tilde f_5 = 0$ for $\sigma =
\sigma_\alpha\simeq0.39696$.

To arrive at equations with a $u_{xxt}$ term, the Helmholtz operator
is applied. The Helmholtz operator fits nicely into the above
procedure, because it is the inverse of the elimination of the
$u_{xxt}$ term, at the order considered.  More precisely, if the
Helmholtz operator $1-f_4\partial_x^2$ is applied to (\ref{step2})
then equation (\ref{step1}) is recovered. Therefore, each of the three
steps leading to the normalised form (\ref{step3}) can also be
inverted, and any two equations with equal $\tilde f_5$ are
asymptotically equivalent. The CH-equation has $\tilde f_5 = {\rm
sign}(\alpha^2 \Gamma)$, even though $f_5 = 0$ originally. Moreover,
$\tilde f_3 = {\rm sign}(\Gamma)$, so that the relative sign is that
of $\alpha^2$.  From these equations, it follows that the vanishing of
$\tilde f_3$ implies the vanishing of $\tilde f_5$. This is not true
in the water wave equation. That is, $\sigma = 1/3$ does not make the
coefficient of $u_{xxxxx}$ vanish. This explains why the CH-equation
is not a good model for $\sigma = 1/3$.

In general, all the terms in the linear function $G$ in equation
(\ref{FG}) may be removed by a Kodama transformation. However, it is
not possible in general at the same time to remove the fifth order
derivative. One may, however, trade the fifth order derivative for the
$u_{xxt}$ term.  Therefore, possibly the simplest representative
equation in the class (\ref{FG}) has the form
\[
   u_t + u_x + uu_x + u_{xxx} + \kappa u_{xxt} = 0 \,,
\]
where small positive/negative $\kappa$ gives the different signs for
$\tilde f_5$. The case with positive $\kappa$ results from the BBM
equation of \textsc{Benjamin {\it{et al.}}} \cite{BBM} by a Galilean transformation and velocity shift
(which is not included in the transformation group
(\ref{txuTraf})). It is known that the BBM equation is not
integrable. Thus, the integrable CH equation arises as a
Kodama-Helmholtz near-identity transformation of the non-integrable
BBM equation, after a Galilean transformation and velocity shift.

\subsection{Examples}

\subsubsection{Asymptotic equivalence of CH
and KdV5 equation}

The previous section constructs a transformation from the
water wave equation to the CH equation. The general argument of this
section shows that it is also possible to transform the CH equation, for
example, into the KdV5 equation.  To this end, we first expand the time
derivative in the
$u_{xxt}$-term using the equation itself and then apply a transformation of
the form
\begin{equation}
    \label{KodamaKdV}
      u =
v + \epsone(\alpha_1v^2+\alpha_2v_x\intx{v})
    + \epstwo\beta v_{xx}\; .
\end{equation}
Choosing the values in this Kodama transformation as
\begin{equation} \label{trafoKdV5}
\alpha_{{1}}={\frac {{\alpha}^{2}}{\Gamma}}
\,, \quad
\alpha_{{2}}=2{\frac {{\alpha}^{2}}{\Gamma}}
\,, \quad
\beta=2{\alpha}^{2}
\,,
\end{equation}
transforms the CH equation (\ref{CH-BV}) into the
integrable KdV5
equation
\begin{equation} \label{KdV5}
v_t+c_0v_x+3vv_x+5\left (vv_{xxx}+2v_xv_{xx}\right
){\alpha}^{2}+
\frac{15}{2}\frac{{\alpha}^{2}{v}^{2}v_x}{\Gamma}
+\Gamma\left( {\alpha}^{2} v_{xxxxx}+v_{xxx} \right) = 0
\,.
\end{equation}

We conclude that (\ref{CH-BV}) is asymptotically equivalent to the
integrable KdV5 equation, and both of them are equivalent to
(\ref{asy1}) at order ${\cal O}(\epstwotwo)$.  However, the
equivalence of (\ref{CH-BV}) to the KdV5 equation breaks down in the
limit $\Gamma \to 0$, because both the transformation and the
resulting equation are singular in the limit $\Gamma \to
0$. Therefore, the peakon equation cannot be transformed into KdV5.

Using the additional parameter $\nu$ supplied by the Helmholtz operator
allows for the removal of the highest order term while preserving the
dispersion relation, which is unchanged by applying a linear operator
to the equation. One advantage of the CH equation over the
asymptotically equivalent KdV5 equation is that it is numerically
easier to integrate because it does not contain the fifth derivative.
This is in accordance with the general smoothing effect of the
Helmholtz operator.

Different variants of higher order KdV equations have been derived in
the literature.In this context, we mention \cite{ChampneysGroves}, who
obtained a 5th order KdV equation by expanding the variational
formulation of the water wave problem. Again, their equation is
asymptotically equivalent to CH by a certain Kodama
transformation. From our point of view, however, the virtue of the
power of the Kodama-transformation is that it allows one to select
integrable equations as the model equation.  The equation derived in
\cite{ChampneysGroves} is not known to be integrable.

\subsubsection{Asymptotic equivalence to the $b$-equation}

Recently a new variant of (\ref{CH-BV}) has been introduced by
\textsc{Degasperis \& Procesi}
\cite{DP[1999]} as
\begin{equation} \label{eqn:CHB3}
      m_t + u m_x + b u_x m = c_0 u_x - \Gamma u_{xxx}
\,,
\end{equation}
where $b$ is an arbitrary parameter. The cases $b=2$ and $b=3$ are special
values for this equation. The case $b=2$ restricts it to the integrable
CH equation. The case $b=3$ is the DP equation of \textsc{Degasperis
\& Procesi} \cite{DP[1999]}, which
was shown to be integrable in \cite{DHH2002}. The two cases CH and DP
exhaust the integrable candidates for (\ref{eqn:CHB3}), as may be
shown using either Painlev\'e analysis, as in \cite{DHH2002}, or the
asymptotic integrability test, as in
\cite{DP[1999]}. The b-family of equations (\ref{eqn:CHB3}) was also shown
in \cite{MN[2002]} to admit the symmetry conditions necessary for
integrability only in the cases
$b=2$ for CH and $b=3$ for DP.

We shall show here that the new integrable DP equation can also be
obtained from the shallow water elevation equation (\ref{asy1}) by an
appropriate Kodama transformation. The derivation in the previous
section is essentially unchanged up to equation (\ref{eqn:twotoone}). The
two scaling relations (\ref{C2},\ref{C3}) now read
\begin{eqnarray*}
(\Aco - \frac92\nu) : (\Bco-\frac32\nu) &=& b:1,
    \\
        \frac32 \nu : (\Bco - \frac32 \nu)    &=&  b+1 :
1 \,.
\end{eqnarray*}
These two conditions imply
\[
\Bco = \nu\frac32 \frac{b+2}{b+1} \quad \hbox{and}
\quad \Aco
= \nu \frac32 \frac{4b+3}{b+1} \,.
\]
The resulting Kodama transformation of the form (\ref{Kodama}) with
coefficients $\alpha_1'$, $\alpha_2'$, and $\beta'$ are
\begin{eqnarray*}
    \alpha_1' &=& \alpha_1 + 3 \Lambda \\ \alpha_2' &=& \alpha_2 -6
    \Lambda\\ \beta' &=& \beta - (1-3\sigma) \Lambda \quad
\mbox{where} \\
\Lambda &=& \frac{b-2}{b+1} \frac{45\sigma^2 +
30\sigma - 19}{360}\,.
\end{eqnarray*}

Therefore any $b \not = -1$ can be achieved by an appropriate Kodama
transformation. Note that when $\sigma =\sigma_\alpha$, see
(\ref{sigmacrit}), then $\alpha^2=0$, hence $\Lambda = 0$ is independent of
$b$.  After this transformation (\ref{eqn:CHB3}) is obtained by further
scaling the new dependent variable $u$ by the factor $b+1$.  See
\cite{HolmStaley2002} for discussions of equation (\ref{eqn:CHB3}) in which
$b$ is treated as a bifurcation parameter when $c_0=0$ and $\Gamma=0$.

The value $b=-1$ is excluded, not from a deficiency of the
KH-transformation, but because the term $uu_x$ is not removable in the
linear function $F$ in equation (\ref{FG}). As we have shown above, the
KH-transformation does not affect this term, and so it cannot be removed
from the $\eta$ equation (\ref{asy1}). That is, the case $b=-1$ is not
within the range of the KH-transformation.

We conclude that the detailed values of the coefficients of the asymptotic
analysis at quadratic order hold only modulo the KH-transformations, and
these transformations may be used to advance the analysis and thereby gain
insight.  Thus, the KH-transformations may provide an answer to
the perennial question ``Why are integrable equations found so often,
when one uses asymptotics in modelling?''

\section{Dispersion relation}
\label{sec:dis}

\begin{figure}
\vspace{0.5cm}
   \centerline{\includegraphics[width=14cm]{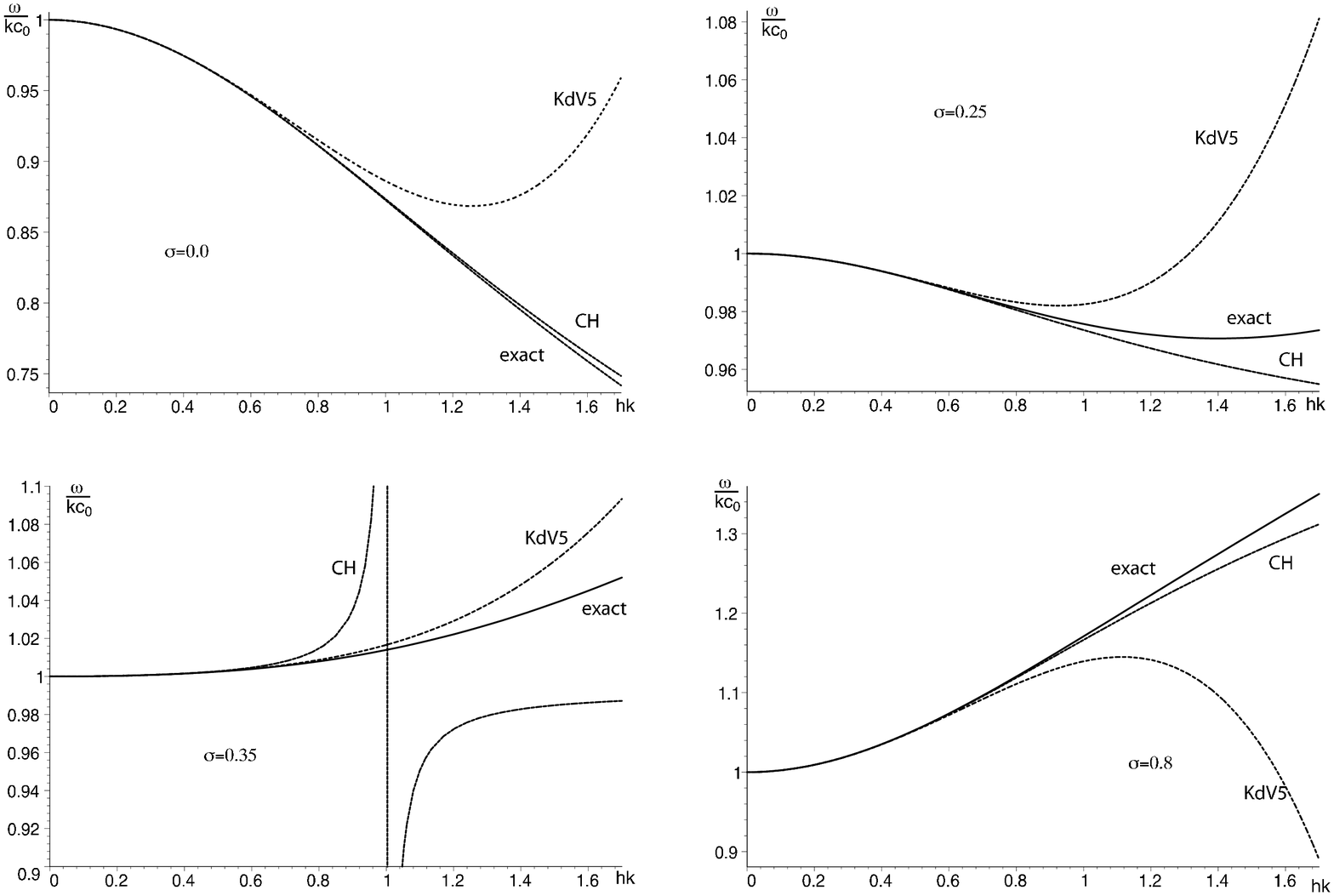}}
\caption{Normalised dispersion relations for
$\sigma = 0.0,0.25, 0.35, 0.8$ as a function of $hk$.
The full line is the exact phase speed (\ref{ch_sw2});
the dashed line is the
approximation (\ref{ch_disp}) \label{fig:disp}
and the short dashed
line is that of KdV5.}
\end{figure}

The interplay between the local and nonlocal linear dispersion in the
CH equation (\ref{peyresq}) is evident in the relation for its phase
velocity,
\begin{equation}
      \frac{\omega}{k} =
             c_0 - \frac{\Gamma k^2}{1 + \alpha^2k^2}\,,
\label{ch_disp}
\end{equation}
for waves with frequency $\omega$ and wave number $k$ linearized
around ${u}=0$.  For $\Gamma<0$, short waves and long waves both travel in
the same direction.  Long waves travel faster than short ones (as
required in shallow water), provided $\Gamma<0$. Then the phase
velocity lies in $\omega/k\in(c_0-\,\Gamma/\alpha^2, c_0]$.
At low wave numbers, the constant dispersion parameters $\alpha^2$ and
$\Gamma$ both perform rather similar functions. At high wave numbers,
however, the parameter $\alpha^2$ properly keeps the phase velocity of
the wave from becoming unbounded, and the dispersion relation is
similar to the original dispersion relation for water waves, provided
that the surface tension vanishes $\sigma=0$.  The remarkably accurate
linear dispersion properties close to $k\approx 0$ give the CH
equation a clear advantage over the KdV equation (provided $\sigma$ is
not close to $1/3$, see Fig. \ref{fig:disp}).

For the peakon equation, $c_0=\Gamma=0$ and linear dispersion is
absent.  For nonvanishing surface tension, the true dispersion
relation for shallow water waves is unbounded for large wave numbers,
whereas the dispersion relation of equation (\ref{CH-BV}) saturates to
the asymptotic value $c_0-\Gamma/\alpha^2$.

Consequently, (\ref{CH-BV}) is inferior to KdV5 for non-zero surface
tension, with respect to traveling wave solutions exhibiting small
tail waves. Unboundedness of the linear dispersion relation of KdV5
for high wave numbers allows resonances to occur between a
supercritical solitary wave and high-wavenumber linear waves. These
resonances give rise to exponentially small ripples at the tails of
the solitary wave, in accord with the water wave solution of the full
Euler equation. See, for example,
\cite{Beale[1991],GrimshawJoshi[1995],DiasKharif[1999]} and
\cite{Lombardi[2000]} for more discussion and analysis of these fascinating
resonances.  Nonetheless, traveling wave solutions of CH and KdV5 are
asymptotically equivalent, and they both agree asymptotically with
traveling wave solutions of the full water wave problem, see sec.
\ref{sec:tw}.

The connections to the physical parameters $\alpha=\alpha(\sigma)$ and
$\Gamma=\Gamma(\sigma)$ are defined in equations (\ref{alpha-gamma}).
The asymptotic equation (\ref{peyresq}) was derived from the water
wave equation by means of the KH-transformation. As explained in
sec.~\ref{subsec:disprel}, the dispersion relation (\ref{ch_disp})
matches the dispersion relation for water waves up to quintic order.
In comparison, the dispersion relation for water waves, when developed
for small wave number $k$ yields
\begin{eqnarray} \label{ch_sw2}
\frac{\omega}{c_0 k}  &=&  \sqrt{\frac{1+\sigma
h^2k^2}{hk} \tanh hk
}        \\
&\approx&   1  - \frac{1}{6} (1-3\sigma) h^2k^2
               + \frac{1}{360}(19 -30\sigma-45\sigma^2)
h^4k^4       \\
&\approx&   1  - \frac{1}{6} (1-3\sigma) h^2k^2
               (1-\nu h^2 k^2) \; .
\end{eqnarray}
Therefore, the dispersion relations are in agreement up to 5th order in
dimensionless wavenumber 
 $hk$.
For the particular value $\sigma \approx 0.069$, the dispersion relations
agree up to 7th order.

\section{Travelling wave solutions}
\label{sec:tw}

One may also ask how the solutions of equation (\ref{CH-BV}) compare
to the usual KdV solitons and to real water solitary waves.  An
expansion of the form of the traveling wave in the full Euler
equations is given by \textsc{Grimshaw} \cite{Grimshaw[1971]}, and to
higher order by \textsc{Fenton} \cite{Fenton[1972]}.
\footnote{For only moderately steep waves, and for the 
corresponding cnoidal waves, Fenton has shown that the 1972 approach
gives a very poor approximation to the velocity field, see Fenton
(1990) [H. Peregrine (private communication)].}
The result given in \cite{Grimshaw[1971]} is normalised so that the
highest point of the wave at $s=0$ is unity.

By direct substitution of a series in $\sech^2$ into (\ref{final}), one finds
\[
      u(s) = \aalpha \sech^2 b s + \frac{19}{20}\epsone\aalpha^2\sech^4 bs \,,
\]
where $b^2 = 3\aalpha\epsone/(4\epstwo)$ and $c = 1 +
\epsone\aalpha/2 + \frac{19}{40}\epsone^2\aalpha^2$ in $s=x-ct$.
Applying the (inverse) Kodama transformation gives
\[
      \eta(s) = \aalpha ( 1 + \frac12 \epsone \aalpha)\sech^2 bs
+ \frac34 \epsone\aalpha^2\sech^4 bs \,.
\]
For simplicity, we restrict the calculation to the case without
surface tension, i.e., $\sigma=0$.  After normalising the height at
the crest to unity, we find perfect agreement up to order $\epsilon
a^2$ with the exact result by \textsc{Grimshaw} \cite{Grimshaw[1971]}.
This shows that the travelling wave solutions of (\ref{CH-BV}) agree
with those of the exact Euler equations up to the order we are
considering, which is the optimal result.  In particular, the
traveling waves are narrower and slower (in this normalisation) than
the KdV soliton, in agreement with experimental findings of
\textsc{Weidman \& Maxworthy} \cite{WeidmanMaxworthy[1978]}. The solution also agrees with the one
found in \cite{LiZhiSibgatullin[1997]} by solving the 5th order KdV
equation.

\section*{ Acknowledgments}

We are grateful to R. Camassa, O. Fringer, I. Gabitov, R. Grimshaw,
Y. Kodama, B.  McCloud, J.  Montaldi, P. J. Olver, T. Ratiu,
M. Roberts, G. Schneider, P. Swart, E. C. Wayne and A. Zenchuk for
their constructive comments
and encouragement. We also thank H. Peregrine for his careful reading
of the manuscript and for his expert comments.
This work began at the MASEES 2000 Summer School in
Peyresq, France.  GAG was supported by a European Commission Grant,
contract number HPRN-CT-2000-00113, for the Research Training Network
{\it Mechanics and Symmetry in Europe\/} (MASIE).  HRD was partially
supported by MASIE.  DDH was supported by DOE, under contract
W-7405-ENG-36.

\end{document}